# Spin-transport characteristics in a Si-based spin metal-oxide-semiconductor field-effect transistor (spin MOSFET): Bias dependence of the spin polarization in Si and magnetoresistance in spin-valve signals

Shoichi Sato[1,2], Masaaki Tanaka[1,2,3], and Ryosho Nakane[1,3,4]

[1]*Department of Electrical Engineering and Information Systems, The University of Tokyo, 7-3-1 Hongo, Bunkyo-ku, Tokyo 113-8656, Japan*
[2]*Center for Spintronics Research Network (CSRN), The University of Tokyo, 7-3-1 Hongo, Bunkyo-ku, Tokyo 113-8656, Japan*
[3]*Institute for Nano Quantum Information Electronics, The University of Tokyo, 4-6-1 Meguro-ku, Tokyo 153-8505, Japan*
[4]*Systems Design Lab (d.lab), The University of Tokyo, 7-3-1 Hongo, Bunkyo-ku, Tokyo 113-8656, Japan*

**Abstract**: We have studied the spin transport characteristics of an enhancement-type spin metal-oxide-semiconductor field-effect transistor (spin MOSFET), particularly the bias voltage dependence of the electron spin polarization $P_S$ in Si and the magnetoresistance ratio $\gamma^{MR}$ in spin-valve signals, to optimize the device performance. The examined spin MOSFET device has an 8-nm-thick $p$-Si channel with a back gate (G) and ferromagnetic source / drain (S/D) junctions consisting of Fe/Mg/MgO/SiO$_x$/$n^+$-Si. In addition to transistor characteristics with an on-off ratio of $10^4$, clear spin-valve signals and Hanle spin precession signals were observed at 4 K in a wide range of the source-to-gate $V_{GS}$ and source-to-drain $V_{DS}$ bias voltages. We achieved a high $P_S$ of 50% and a high $\gamma^{MR}$ of 0.35% as the maximum values in their single-peaked curves plotted as a function of the junction voltage $V_J$, mainly because the ferromagnetic S/D junction can generate high $P_S$ and the spin diffusion length is very long in the Si inversion channel. These $P_S$ and $\gamma^{MR}$ values are the highest ever reported in spin-MOSFETs. Our spin transport model for our spin MOSFET structure was improved in this study by taking into account the electron distribution and band profile of the $n^+$-Si regions in the ferromagnetic S/D junctions, which enables the accurate estimation of $P_S$. Detailed analyses with various $V_{GS}$ and $V_J$ clarified that $P_S$ is determined only by $V_J$. Our analyses also revealed that the main parameters for determining $\gamma^{MR}$, such as $P_S$ and the resistance-area products of the S/D ferromagnetic junctions, have different $V_J$ dependences, leading to the finding that the present device does not exploit the full potential of the ferromagnetic S/D junctions to maximize $\gamma^{MR}$. Based on the results, we discuss the device physics and engineering for further enhancement of $\gamma^{MR}$, with a focus on the electrical and spin-related properties of the ferromagnetic S/D junctions.

# I. Introduction

Si-based spin metal-oxide-semiconductor field-effect-transistors (spin MOSFETs) have been actively studied during the last two decades [1-12]. This is because spin-MOSFETs have a high potential for versatile applications in next-generation internet-of-things (IoT), such as low-power digital/analog circuits and high-density nonvolatile memory, using their reconfigurable transistor characteristics obtained by controlling the relative magnetization direction between ferromagnetic source (S) and drain (D) electrodes. To fully take the technical advantages of spin MOSFETs, it is indispensable to realize not only a high on-off ratio of the transistor action, but also a high magnetoresistance ratio $\gamma^{MR}$ that is defined by the maximum change in the device (D-S) resistance between the parallel (P) and antiparallel (AP) magnetization configurations. So far, several groups have reported fundamental operations of spin MOSFETs at room temperature; however, sufficiently high $\gamma^{MR}$ for practical use has yet to be achieved: The maximum $\gamma^{MR}$ value [13,14] is lower than 0.04% at 300 K [6] and 0.12% at 3 K [5].

In our previous paper [9], we fabricated a bottom-gate-type spin MOSFET on a silicon-on-insulator (SOI) substrate, using a lightly-doped $p$-Si layer for a Si two-dimensional (2D) inversion channel and Fe/Mg/MgO/$n^+$-Si ferromagnetic tunnel junctions for the S and D electrodes. We demonstrated device operations at 295 K; the on-off ratio is ~$10^6$ and the spin polarization $P_S$ of an electron current is almost unchanged when electrons travel from the S to the D electrode through the Si 2D channel, i.e., the spin transport efficiency estimated from the difference in $P_S$ between the S and D reaches 94% with the aid of the spin drift effect [20-24]. Despite such high capability of the Si 2D channel, $\gamma^{MR}$ was only ~0.02%. Detailed analyses using our spin transport model revealed that the main origins of such low $\gamma^{MR}$ are insufficient $P_S$ (= 5 − 7% at 295 K and 14% at 4 K [9]) and large resistance-area product $R_A$ (~5000 Ωμm$^2$ at 295 K and ~7000 Ωμm$^2$ at 4 K [15]) of the S and D junctions.

To achieve higher $\gamma^{MR}$, Fe/Mg/MgO/SiO$_x$/$n^+$-Si junctions are very attractive because our previous work using three-terminal (3T) vertical devices fabricated on a $n^+$-Si substrate demonstrated that they can generate a high $P_S$ up to 42% in $n^+$-Si owing to a low electron trap density at the SiO$_x$/$n^+$-Si interface [15]. Furthermore, we studied the spin transport via a Fe/Mg/MgO/SiO$_x$/$n^+$-Si junction using a four-terminal lateral device with a $n^+$-Si channel on a SOI substrate, and clarified the mechanism of the spin-selective tunneling at 4 K [14]. Our spin transport model taking into account the band diagram of the junction well explains almost all the features of the experimental spin transport signals measured in a wide range of the junction voltage $V_J$. Our prominent finding is that $V_J$ is a key parameter to obtain high $P_S$ in both the spin injection and extraction geometries since $P_S$ has a strong $V_J$ dependence. From these previous results, when the Fe/Mg/MgO/SiO$_x$/$n^+$-Si junction structure is used for the S and D in a spin MOSFET, $\gamma^{MR}$ is expected to be enhanced; however, it is unclear whether the device can use the full potential of the junction capability. This is because the device characteristics arising from the junction properties, such as the junction band diagram, can largely change from those of our previous spin MOSFETs [9] with different junction structure without SiO$_x$ (Fe/Mg/MgO/$n^+$-Si). Thus, it is necessary to systematically study how $\gamma^{MR}$, $P_S$, and other parameters in the spin transport change in relation to each other when the bias condition is varied in a wide range. Moreover, if spin transport signals are greatly enhanced, other characteristics in the previous studies, such as slight change in $P_S$ depending on $V_J$ and $V_{GS}$ (shown in Fig. S8(a) in Supplemental Material (S.M.) [17]), may become clear, which can lead to further improvement in our spin transport model.

In this paper, we study the spin transport characteristics of a spin MOSFET with Fe/Mg/MgO/SiO$_x$/$n^+$-Si junctions for the S and D electrodes with various drain-source voltages $V_{DS}$ and gate-source voltages $V_{GS}$, towards the enhancement of $\gamma^{MR}$. In particular, critical parameters in the spin transport, such as $P_S$ and $\gamma^{MR}$, and the band diagrams in the S and D junctions are thoroughly investigated under various bias conditions using our spin transport model. Finally, the strategy for further enhancement of $\gamma^{MR}$ is discussed in terms of the ferromagnetic tunnel junction properties.

**II. Device preparation**

Figure 1(a) shows a schematic device structure of our spin MOSFET examined in this study, which was fabricated on a SOI substrate with a 200-nm-thick SiO$_2$ buried oxide (BOX) layer and a 8-nm-thick $p$-Si channel with a boron doping concentration of $N_A = 1\times10^{15}$ cm$^{-3}$. The Cartesian coordinate is defined as follows: the $x$ and $y$ axes are parallel to the channel width and length directions, respectively, and the $z$ axis is perpendicular to the substrate plane. In this device, Fe/Mg/MgO/SiO$_x$/$n^+$-Si junctions are used as the S and D electrodes, the BOX layer is used as the gate dielectric, and reference electrodes L and R are located outside of the S and D electrodes, respectively. The fabrication process is basically the same as that in our previous paper [9]. First, a 120-nm-thick SiO$_2$ layer was formed on a $p$-Si (SOI) layer by thermal oxidation, and it was etched with buffered HF (BHF) to open the Si surface of the contact areas. These areas were doped with phosphorous (P) donor atoms through thermal diffusion from a P$_2$O$_5$ film on the Si surface, and $n^+$-Si regions (donor concentration $N_D = \sim 10^{20}$ cm$^{-3}$) shown by the yellow regions in Figs. 1(a) and (b) were formed. The $n^+$-Si regions are introduced to reduce the parasitic resistances arising from the Schottky barrier formed under the S/D electrodes. After chemical etching of the P$_2$O$_5$ film and successive surface cleaning of the $n^+$-Si regions, the substrate was installed into an ultrahigh vacuum system and the surface of the $n^+$-Si region was thermally cleaned. An MgO layer (light brown region in Fig. 1(b)) was deposited on the $n^+$-Si region by electron beam evaporation at room temperature. Next, a $SiO_x$ layer was formed at the interface between MgO and $n^+$-Si by radio frequency (RF) plasma oxidation [15], in which the substrate temperature was set at room temperature, the oxidation time was 1 min, and the RF power and $O_2$ pressure were 100 W and $2 \times 10^{-3}$ Pa, respectively. The thickness of the $SiO_x$ layer was estimated to be 0.2 nm from transmission electron microscopy (TEM) observation [15]. On the MgO surface, (from top to bottom) an Al (15 nm)/Mg (1 nm)/Fe (4 nm)/Mg (1 nm) layered structure was fabricated through a successive deposition at room temperature by molecular beam epitaxy (MBE). After being exposed to air, a back-gate-type MOSFET structure was fabricated using the same procedure as in Ref. [9]. A detailed cross-sectional view of the S/D junctions and the channel are schematically illustrated in Fig. 1(b), in which pink, dark brown, light brown, green, yellow, and light blue regions are Fe, Mg, MgO, $SiO_x$, $n^+$-Si, and $p$-Si, respectively, the lengths of the S and D along the $y$ direction are $L_S = 0.7$ μm and $L_D = 2.0$ μm, respectively, the designed channel length $L_{ch}$ is 0.5 μm, and the width of the electrodes and channel is $W_{ch} = 180$ μm. The effective channel length $L_{eff} = L_{ch} + \Delta L$ in Fig. 1(b) will be electrically estimated for precise analysis of the spin transport. Note that the electrodes L and R are not used in this study. We also prepared a Hall-bar-type MOSFET having the same channel structure and properties as those in the spin MOSFET on the same wafer, to characterize the electrical properties of the Si 2D channel, such as the channel sheet resistance $R_S$, sheet carrier density $N_S$, and effective electron mobility $\mu_e$ (these are shown in Section S1 in S.M. [17]).

**III. Experimental results**

Figures 2(a) and (b) show drain-source current $I_{DS} - V_{DS}$ and $I_{DS} - V_{GS}$ characteristics, respectively, measured at 4 K, with the measurement setup in the inset of (a). In Fig. 2, the device exhibits (a) good transistor characteristics and (b) a high on-off ratio of ~$10^4$, and the downward curvature of $I_{DS}$ clearly seen in the range of $V_{GS} \geq 30$ V in Fig. 2(a) is due to non-linear current-voltage characteristics of the MgO/$SiO_x$ tunnel junctions, which indicates that the tunnel junction resistance $R_J$ is larger than the channel resistance $R_{ch}$ when $V_{GS} \geq 30$ V. Two-terminal (2T) spin-valve signals were measured at 4 K using the setup in Fig. 1(a), in which a constant current $I_{DS}$ is driven from the D to S electrode and the change in $V_{DS}$ (= $\Delta V_{DS}$) is measured with a voltage meter illustrated by "$V_{DS}$" while the relative magnetization direction between the S and D electrodes is changed from P to AP and vice versa, by the application of a sweeping in-plane magnetic field $H_{//}$ along the $x$ direction. In our setup, the positive $I_{DS}$ corresponds to the situation that electron spins are injected from the S electrode into the Si 2D channel and detected by the D electrode, whereas the negative $I_{DS} < 0$ corresponds to the situation that electron spins are injected from the D electrode and detected by the S electrode. Figure 2(c) shows $\Delta V_{DS}$ at various $I_{DS}$ values (= −4 – 4 mA) and at $V_{GS}$ = 40 V, where red and blue curves are the positive and negative sweep directions of $H_{//}$, respectively. Clear two-terminal (2T) spin-valve signals indicate spin injection, transport, and detection in the spin MOSFET. Similar 2T spin-valve signals with various $I_{DS}$ values were also obtained at $V_{GS}$ = 60 and 80 V (see Section S2 in S.M. [17]). Then, the amplitude of the spin-valve signal $\Delta V_{SV}$ defined in Fig. 2(c) was estimated for all the $I_{DS}$ and $V_{GS}$ conditions. Figure 2(d) shows $\Delta V_{SV}$ plotted as a function of $I_{DS}$, where blue circles, green squares, and orange diamonds are those estimated at $V_{GS}$ = 40, 60, and 80 V, respectively. We found that $\Delta V_{SV}$ increases with decreasing $V_{GS}$, while the saturation trend of $\Delta V_{SV} - I_{DS}$ characteristics at $|I_{DS}| > 2$ mA becomes noticeable as $V_{GS}$ is decreased. The latter saturation feature at $V_{GS}$ = 40 and 60 V is inconsistent with the basic spin-drift transport theory: $|\Delta V_{SV}|$ increases with the increase of its change rate as $|I_{DS}|$ is increased, owing to the positive increase of the effective spin transport length with $|I_{DS}|$ [20-24]. Our previous paper [9] demonstrated that a spin MOSFET with the same channel structure and lower $R_A$ of the junction has the $\Delta V_{SV} - I_{DS}$ relationship fully following such spin-drift theory. Since there is no difference in the channel properties between the present and previous devices, the features at $V_{GS}$ = 40 and 60 V in Fig. 2(d) come from the decrease in $P_S$ as the junction voltage $V_J$ increases [14], which is more pronounced when $V_J$ becomes comparable to the voltage drop of the Si channel. This consideration about the decrease in $P_S$ is also supported by our previous paper [15]: In three-terminal (3T) devices having the same ferromagnetic junction structure with a high $R_A$ (= ~2.5 ×10$^5$ Ωμm$^2$), their $P_S$ values estimated from the 3T Hanle precession signals monotonically decrease with increasing the constant bias current through the junction.

To estimate various physical parameters, including $\Delta L$ in Fig. 1(b), 2T Hanle precession signals were measured at 4 K using the setup in Fig. 1(a) [10,18,19]. First, the P (or AP) magnetization configuration was prepared using the major (or minor) loop by sweeping $H_{//}$, and then the voltage change $\Delta V_{DS}$ was measured at a constant $I_{DS}$ (= 2 mA) while a sweeping magnetic field was applied perpendicular to the plane ($H = H_\perp$). Figure 3(a) shows $\Delta V_{DS}$ signals obtained at $V_{GS}$ = 40, 60, and 80 V, where red and blue curves are $\Delta V_{DS}^P$ and $\Delta V_{DS}^{AP}$ that correspond to $\Delta V_{SV}$ in the P and AP, respectively. Then, a 2T Hanle precession signal $\Delta V_{2TH}$ was extracted by $\Delta V_{2TH}^{AP(P)}(H_\perp) = +(-)\left[\Delta V_{DS}^{AP}(H_\perp) - \Delta V_{DS}^{P}(H_\perp)\right]/2$, where $\Delta V_{2TH}^P$ and $\Delta V_{2TH}^{AP}$ correspond to $\Delta V_{2TH}$ in the P and AP, respectively. Figure 3 (b) shows $\Delta V_{2\text{TH}}$ at $V_{\text{GS}}$ = 40, 60, and 80 V, where blue and red curves are $\Delta V_{2\text{TH}}^{\text{P}}$ and $\Delta V_{2\text{TH}}^{\text{AP}}$ respectively. The difference between $\Delta V_{2\text{TH}}^{\text{P}}$ and $\Delta V_{2\text{TH}}^{\text{AP}}$ at $H_\perp$ = 0 Oe is consistent with the maximum change in $\Delta V_{\text{SV}}$ under the same bias condition, which ensures that we detected the spin precession signal via the Si 2D channel. The 2T Hanle signals do not exhibit clear oscillating features that were seen for the spin MOSFET with $L_{\text{ch}}$ = 10 μm in our previous paper [9]. This is because the dephasing of the spin precession arising from the spin diffusion is more significant than the in-phase spin precession assisted by the spin drift effect, owing to the shorter channel length ($L_{\text{ch}}$ = 0.5 μm) than the spin diffusion length (~1 μm) [11].

**III. Analysis of spin polarization under various bias conditions**

The 2T spin-valve and Hanle signals obtained in the previous section are analyzed to estimate the $P_{\text{S}}$ values under various $I_{\text{DS}}$ and $V_{\text{GS}}$ conditions. In our previous study [9], we proposed a spin-transport model, in which we neglected a realistic feature that the electron density $n(z)$ in the $n^+$-Si region is distributed along the vertical $z$-direction. In other words, we simply assumed that the electron density in the $n^+$-Si region is uniform and identical with the donor concentration $N_{\text{D}}$ of $1\times10^{20}$ cm$^{-3}$ that was estimated by a Hall measurement of the reference sample with a $n^+$-Si channel (see Section S1 in S.M. [17]). We found that our previous spin transport model cannot well explain the slightly varying $P_{\text{S}}$ with $V_{\text{GS}}$ in our previous study [9] (see Fig. S8(a) in S.M. [17]). Moreover, when we analyze the *present* data in Fig. 2(d) by our *previous* model, $P_{\text{S}}$ exceeds 100% and it varies greatly depending on $V_{\text{GS}}$ (see Fig. S12 in S.M. [17]). This is probably because the change in $n(z)$ with $V_{\text{GS}}$ and $V_{\text{J}}$ was not taken into account in our previous model. This consideration is supported by the fact that $P_{\text{S}}$ is inversely proportional to the spin resistance $r_n$ of the $n^+$-Si region beneath the S and D electrodes [9]; this means that $P_{\text{S}}$ is overestimated when $r_n$ is underestimated.

Here, we reconstruct our spin transport model by taking into account $n(z)$ along the $z$ direction in the $n^+$-Si region. Then, $P_{\text{S}}$ is estimated under the condition that $P_{\text{S}}$ depends only on $V_{\text{J}}$ but not on $V_{\text{GS}}$. This consideration is valid for the following reasons: Based on the spin-dependent tunneling model [14], $P_{\text{S}}$ is determined by the spin polarization of the parabolic band structure of Fe and the differential tunnel conductance ($dJ_e/dV_{\text{J}}$ where $J_e$ is the tunnel current density), thus $P_{\text{S}}$ depends only on $V_{\text{J}}$. In addition, the electric field induced by $V_{\text{GS}}$ applied from the bottom side is almost screened by electrons in the $n^+$-Si regions, which is verified by an electro-static simulation: $V_{\text{J}}$ slightly increases as $V_{\text{GS}}$ is increased, however, its change is less than 1 mV when $V_{\text{GS}}$ is increased from 0 to 100 V at $N_{\text{D}} > 3\times10^{18}$ cm$^{-3}$ (see Fig. S6 in S.M. [17]).

**III-A Reconstruction of the spin transport model**

First, $V_J$, which is the critical parameter that determines $P_S$, was estimated from the data in Fig. 2(a). Here, we define $V_J$ as the electric potential difference between the Fe/Mg electrode and $n^+$-Si region, i.e., the voltage drop in the MgO/SiO$_x$ insulating bilayer. Since $|V_J|$ is far larger than the voltage drop inside the $n^+$-Si region, it is reasonable to approximate $V_J = (V_{DS} - I_{DS} \times R_{ch})/2$, where $V_{DS}$ and $I_{DS}$ are shown in Fig. 2(a) and $R_{ch}$ was estimated from a Hall measurement (see Section S1 in S.M. [17]). In Fig. 4 (a), $V_J$ is plotted as a function of $I_{DS}$, where blue, green, and orange curves are $V_J$ at $V_{GS} = 40$, 60, and 80 V, respectively. The $V_J - I_{DS}$ relations for all the $V_{GS}$ values are similar to each other, and they slightly decrease as $V_{GS}$ is increased. (Note that, as shown in Fig. S6 in S.M., $V_J$ slightly increases as $V_{GS}$ is increased in the electro-static simulation. This disagreement is probably due to the oversimplification in the simulation. For example, we assumed uniformity in the $x$- and $y$-directions in the $n^+$-Si regions and neglected the electric field concentration at the edge of the electrode, which can be modulated by $V_{GS}$.)

Next, we reconstruct our model in the previous paper [9] by incorporating $n(z)$ that has a great influence on the estimation of $P_S$ from the 2T spin-valve signals. Figure 4(b) shows a one-dimensional potential distribution of the conduction band minimum $E_C(z)$ along the vertical $z$ direction in the $n^+$-Si region, where the origin of the $z$ axis is set at the SiO$_x$/Si interface and $\Phi_{pin}$ is the band offset arising from the Fermi level pinning at the SiO$_x$/Si interface [25-27]. The Fermi level pinning will be verified in the later section. The band bending occurs near the Si/BOX interface by the application of $V_{GS}$. Since the $n^+$-Si thickness $t_n = 5$ nm is much thinner than the spin diffusion length (~ 1 μm) [9], the chemical potentials for up and down spin electrons, which are red and blue horizontal lines in Fig. 4(b), are regarded as uniform along the $z$ direction. Since the electron transport in the $n^+$-Si region is dominated by the Coulomb scattering due to the high $N_D$ value of ~$10^{20}$ cm$^{-3}$, it is reasonable to assume that both the electron mobility $\mu_n$ and spin lifetime $\tau_n$ remain unchanged while $V_{GS}$ is varied. Based on this setting, the spin resistance $r_n$ (Ω) in the $n^+$-Si region is expressed as follows (detailed derivation is described in Section S4 in S.M. [17]):

$$r_n(V_{GS}, V_J) = \frac{\lambda_n}{W_{ch} q \mu_n N_S^n(V_{GS}, V_J)}, \tag{1}$$

$$\lambda_n = \sqrt{D_n \tau_n}, \tag{2}$$

where the sheet electron density $N_S^n(V_{GS}, V_J)$ (cm$^{-2}$) is expressed by

$$N_S^n(V_{GS}, -(+)V_J) = N_{S0}^n + C_{BOX}(V_{GS} - V_{th})/q - (+)\alpha C_J V_J / q, \tag{3}$$

$$N_{S0}^n = \int_0^{t_n} n(z)\big|_{V_{GS}-V_{th}=0, V_J=0} dz, \tag{4}$$

where $\lambda_n$ is the spin diffusion length in the $n^+$-Si region, $D_n$ is the electron diffusion constant, $N_{S0}^n$ (cm$^{-2}$) is the initial sheet electron density in the $n^+$-Si region without the electrons induced by the $V_{GS}$ or $V_J$ application (corresponding to the one without band bending in the Si inversion channel), $C_{BOX}$ (F/cm$^2$) is the gate capacitance, $V_{th}$ is the threshold voltage of the Si 2D channel, $q$ is the elementary charge, and $C_J$ (F/cm$^2$) is the capacitance of the MgO/SiO$_x$ bilayer. The sign − (+) in Eq. (3) expresses the depletion (accumulation) of the electrons at the S (D) electrode when $V_J > 0$ is applied. Nondimensional $\alpha$ (= 0 – 1) is the efficiency of the electron accumulation induced by $V_J$, which is determined by the Fermi level pinning at the Si surface: $\alpha = 0$ corresponds to the case that the Fermi level is strongly pinned at a specific energy, whereas $\alpha = 1$ corresponds to the case that the Fermi level is free. Equation (1) means that $r_n$ at the S and D electrodes has different dependences on $V_{GS}$ and $V_J$ through Eq. (3): $r_n$ at the D electrode decreases as the increase in $V_{GS}$ and/or $V_J$, whereas $r_n$ at the S electrode decreases as $V_{GS}$ is increased and/or $V_J$ is decreased. This feature becomes more prominent as $N_{S0}^n$ becomes smaller. Figure 4(c) shows $r_n$ calculated using Eqs. (1)–(3) with various $N_{S0}^n$ values and $V_J = 0$, which are independent of $\alpha$ since the third term in Eq. (3) is zero at $V_J = 0$. When $N_{S0}^n$ is large (~ 5×10$^{13}$ cm$^{-2}$), $r_n$ becomes small (~ 4 Ω) and it is almost independent of $V_{GS}$ since its change is less than 1 Ω when $V_{GS}$ is 20 – 100 V. On the other hand, when $N_{S0}^n$ is small (~ 1×10$^{12}$ cm$^{-2}$), $r_n$ becomes larger and it steeply increases as $V_{GS}$ is decreased. The $N_{S0}^n$ and $\alpha$ values are determined by $N_D$ and the strength of the pinning at the MgO/SiO$_x$/Si interface; however, they cannot be measured directly. Thus, in this paper, these values are estimated through the analysis on the spin-valve signals, as will be described in Section III-C.

The analysis of the 2T Hanle signals in Fig. 3(b) was performed to validate the reconstructed model as well as to estimate the effective channel length $L_{eff} = L_{ch} + \Delta L$. For this purpose, the fitting function is modified from Eq. (4a) in Ref. [9] to Eq. (5) as shown below in the following manner; $r_n$ in Eqs. (5c) and (5d) in Ref. [9] is replaced with $r_n^{S(D)} = r_n(V_{GS}, -(+)V_J)/\sqrt{1+i\gamma_e H_\perp \tau_n}$ using $r_n$ in Eq. (1) where $\gamma_e = 1.76\times 10^7$ Oe$^{-1}$s$^{-1}$ is the geomagnetic ratio of electrons, and $L_{ch}$ in Eq. (4c) in Ref. [9] is replaced with $L_{eff} = L_{ch} + \Delta L$ [10]. In consequence, the modified fitting function is expressed as follows;

$$\Delta V_D^{2TH(P/AP)}(H_\perp) = -\sigma^{P/AP}\,\text{Re}\left[\frac{P_S^2 I_{DS}}{X}\left(\frac{1}{r_{ch}^u}+\frac{1}{r_{ch}^d}\right)\exp\left(-\frac{L_{eff}}{\hat{\lambda}_{ch}^d}\right)\right], \tag{5}$$

$$X = \left(\frac{1}{r_{NL}^{(S)}} + \frac{1}{r_{ch}^d}\right)\left(\frac{1}{r_{NL}^{(D)}} + \frac{1}{r_{ch}^u}\right) - \left(\frac{1}{r_{NL}^{(S)}} - \frac{1}{r_{ch}^u}\right)\left(\frac{1}{r_{NL}^{(D)}} - \frac{1}{r_{ch}^d}\right)\exp\left(-\frac{L_{eff}}{\hat{\lambda}_{ch}^u} - \frac{L_{eff}}{\hat{\lambda}_{ch}^d}\right), \quad (6)$$

$$r_{ch}^{u(d)} = \frac{R_S \hat{\lambda}_{ch}^{d(u)}}{W_{ch}}, \quad r_{ch} = \frac{R_S \hat{\lambda}_{ch}}{W_{ch}}, \quad r_{NL}^{S(D)} = \frac{r_{ch} + r_n^{S(D)} \tanh(L_{S(D)}/\hat{\lambda}_n)}{r_n^{S(D)} + r_{ch} \tanh(L_{S(D)}/\hat{\lambda}_n)} r_n^{S(D)}, \quad (7)$$

where $\hat{\lambda}_{ch} = \sqrt{D_e \tau_S/(1+i\gamma H_\perp \tau_S)}$ and $\hat{\lambda}_n = \sqrt{D_n \tau_n/(1+i\gamma H_\perp \tau_n)}$ are the complex spin diffusion length in the Si 2D channel and $n^+$-Si region, respectively, $\hat{\lambda}_{ch}^{u(d)} = \left[+(-)v_d/2D_{ch} + \sqrt{(v_d/2D_{ch})^2 + \hat{\lambda}_{ch}^{-2}}\right]^{-1}$ is the complex up (down)-stream spin diffusion length in the Si 2D channel, and $v_d$, $\tau_S$, and $D_{ch}$ are the electron drift velocity, spin lifetime, and diffusion constant in the Si 2D channel, respectively (see Section S5 in S.M. [17] for details). Fitting parameters $\Delta L$, $P_S$, $N^n_{S0}$, and $\alpha$ are estimated while the other parameters listed in Section S1 in S.M. [17] are fixed at the values estimated in our previous study [11]. It is worth noting that $\Delta L$ is independently estimated from the shape of the 2T Hanle signal, whereas $P_S$ is estimated after $N^n_{S0}$ and $\alpha$ are set at specific values. Thus, first the fitting is performed to estimate $\Delta L$, and then $P_S$, $N^n_{S0}$, and $\alpha$ are estimated in Section III-B, as described below. The fitting results shown by the dashed curves in Fig. 3(b) agree very well with the experimental 2T Hanle signals. For all the $V_{GS}$ values, $\Delta L$ was estimated to be 0.36 μm that reasonably agrees with 0.5 μm estimated in our previous paper [10]. These agreements are convincing evidence that our reconstructed model can be applied to the analysis of the Hanle signals.

### III-B Determination of $N^n_{S0}$ and $\alpha$ by regression analysis

In this section, $N^n_{S0}$ and $\alpha$ are estimated from the spin-valve signals $\Delta V_{SV}$ measured with various $V_J$ and $V_{GS}$ in Fig. 2(d), based on the consideration that $P_S$ depends only on $V_J$, but not on $V_{GS}$, as explained earlier. This consideration is equivalent to that the $P_S$–$V_J$ plot with different $V_{GS}$ match each other when the optimum $N^n_{S0}$ and $\alpha$ values are chosen. Our estimation procedure is as follows: First, $P_S$ is calculated from $\Delta V_{SV}$ in Fig. 2(d) using Eq. (5) ($\Delta V_{SV} = \Delta V_D^{2TH(AP)}(0) - \Delta V_D^{2TH(P)}(0)$) with some specific $N^n_{S0}$ and $\alpha$ values, and then the obtained $P_S$ is plotted as a function of $V_J$. Next, a regression analysis (described below in more detail) was used to evaluate the similarity among the $P_S$–$V_J$ plots with different $V_{GS}$. Finally, the optimum $N^n_{S0}$ and $\alpha$ values are explored and determined so that the above-mentioned similarity is maximized (all the $P_S$–$V_J$ plots match each other) while $N^n_{S0}$ and $\alpha$ were varied in a wide range.

As demonstrated in our previous paper [14], $P_S$ for certain $V_J$ values can be calculated using our spin transport model with the band diagram of a junction in a self-consistent manner. On the other hand, the $\Delta V_{DS}$ measurements under a sweeting magnetic field at a certain $I_{DS}$ value lead to different $V_J$ values when $V_{GS}$ is varied. In this way, $P_S$–$V_J$ plots are obtained as shown in Fig. 4(d). Here, we cannot analyze the $P_S$–$V_J$ curve in a unified manner using a physics-based function while just changing $V_J$ as a parameter. For this reason, we use a polynomial regression with a 4th order polynomial function,

$$P_S^{model}(V_J) = \sum_{i=0}^{4} a_i V_J^i$$

where $a_i$ ($i = 0 – 4$) is constant, to determine the optimum $N_{S0}^n$ and $\alpha$. The 4th order polynomial is used as a model function since this order is the minimum requirement to reproduce the $P_S$–$V_J$ curve in $V_J > 0$, which is theoretically expected to have an asymmetric peak shape [14] (shown by the red solid curve in Fig. S7 in S.M. [17]). In the optimization process by the regression, the coefficient of determination $\Gamma^2 = 1 - \sum_{for\ all\ V_J}(P_S - P_S^{model})^2 / \sum_{for\ all\ V_J}(P_S - <P_S>)^2$ (where $<P_S>$ is the average of $P_S$) between a model polynomial function $P_S^{model}$ and experimental data $P_S$ is used as a quantitative index to measure their similarity; the required condition is satisfied as $\Gamma^2$ is closer to 1. After several explorations, the condition that $N_{S0}^n = 2 \times 10^{12}$ cm$^{-2}$ and $\alpha = 0$ was found to have the highest $\Gamma^2 \sim 0.95$. The same values were obtained from the regression analysis using the $\Delta V_{SV}$ data in $V_J < 0$ in Fig. 2(d). The $N_{S0}^n$ value (= $2 \times 10^{12}$ cm$^{-2}$) is much smaller than the designed value ($N_D \times t_n \sim 5 \times 10^{13}$ cm$^{-2}$); the origin of this small $N_{S0}^n$ is the Fermi level pinning at the SiO$_x$/Si interface and band bending in the n$^+$-Si region, as discussed in Section III-C. The $\alpha$ value (= 0) indicates the presence of the Fermi level pinning at the SiO$_x$/n$^+$-Si interface.

Figure 4 (d) shows $P_S$ estimated using $\Delta L = 0.36$ μm, $N_{S0}^n = 2 \times 10^{12}$ cm$^{-2}$, and $\alpha = 0$, plotted as a function of $V_J$, where blue circles, green squares, and orange diamonds correspond to $V_{GS} = 40$, 60, and 80 V, respectively, like those in Fig. 2(d). The $P_S$–$V_J$ curves with three $V_{GS}$ values are well matched and they are symmetric with a peak in each positive and negative $V_J$ range. The maximum peak $P_S$ value is 50 % at $V_J = 0.33$ V in $V_J > 0$. These features are consistent with our previous results: The $P_S$ values are comparable to those estimated by the 3T method using the same junction structure ($P_S = 34 – 41\%$ at $V_J = 0.6 – 0.9$ V) [15], and the peaked feature seen in the $P_S$–$V_J$ curves agrees well with our theoretical analysis [14] (see Section S6 in S.M. [17]). Note that $P_S$ ($\sim 50\%$) at the small $V_J \sim \pm 0.3$ V in Fig. 4(d) exceeds the spin polarization of Fe ($P_{Fe} \sim 42\%$ [31]), which mainly comes from the enhancement in $P_{det}$ by the small extraction bias application: As shown in Fig. 8(b) in Ref. [14], $P_{det}$ at $V_J \sim -0.3$ V is roughly three times larger than $P_{det}$ at $V_J = 0$ V, leading to ~1.7 times enhancement in $P_S$ (= $(P_{inj}P_{det})^{0.5}$), where $P_{inj}$ is spin injection polarization and $P_{det}$ is spin detection polarization. This enhancement originates from the non-linear $I$–$V$ characteristics expressed by Eq. (16) in Ref. [14] (the detailed explanation is described in Section S9 in S.M. of ref. [14]). On the other hand, $P_S$ at $V_J \sim 0$ V in our device is ~40% as extrapolated from the $P_S$–$V_J$ plot in Fig. 4(d). This is also consistent with the theoretical prediction; $P_S \sim P_{Fe} \sim 40$ % at $V_J \sim 0$ V is expected since the linear approximation is valid when $V_J \sim 0$ V, i.e., the situation corresponds to that predicted by Valet and Fert [34]. Therefore, the $P_S$ values in Fig. 4(d) are reasonable. Based on the theory in Ref. [14], the peaked feature in the $P_S$–$V_J$ curves originates from the competition of the following two changes caused by the increase in $V_J$; $P_{det}$ steeply increases arising from the non-linear $I$–$V$ characteristics as explained above, and $P_{det}$ monotonically decreases because the spin polarization of Fe decreases with increasing the electron energy (the spin-split parabolic band structure has lower spin polarization in the higher energy states as shown in Fig. 10 in Ref. [14]).

Based on the $P_S$–$V_J$ trends in Fig. 4 (d), the characteristic dependence of $\Delta V_{SV}$ on $I_{DS}$ ($V_J$) and $V_{GS}$ in Fig. 2(d) can be explained as follows. As $V_{GS}$ is increased, $\Delta V_{SV}$ decreases through the decrease in $r_n$ in Fig. 4(c). As $I_{DS}(V_J)$ is increased, the increase rate of $\Delta V_{SV}$ decreases through the decrease in $P_S$ in Fig. 4(d).

**III-C Origin of the electron density reduction**

The remaining question is why the estimated $N_{S0}^n$ (= $2 \times 10^{12}$ cm$^{-2}$) value is much smaller than the designed value (~$5 \times 10^{13}$ cm$^{-2}$). Based on the analyses so far, this result probably originates from the Fermi level pinning at the SiO$_x$/$n^+$-Si interface. To examine this possibility, we have to clarify the electron density distribution $n(z)$, because $N_{S0}^n$ in Eq. (3) was varied as a parameter to satisfy the required condition in the regression analysis as described earlier. Here, the potential $E_C(z)$ and $n(z)$ in the $n^+$-Si region are calculated by the Poisson's equation and a classical electron distribution. The boundary condition at the backside channel is the continuity of the electric flux that was estimated in our previous paper [9]. An assumption is that $N_D$ is in the range of degenerated Si (> $3 \times 10^{18}$ cm$^{-3}$) and $\Phi_{pin}$ is located in the band gap of Si ($\Phi_{pin} = 0.82 \pm 0.36$ eV) [25-29]. The former is supported by the fact that the $I$–$V$ curve does not have a rectifying characteristic, i.e., the Schottky barrier is thin enough owing to the high $N_D$ (> $3 \times 10^{18}$ cm$^{-3}$). The latter is supported by our previous paper [30], in which the *I–V* curve is almost identical with those obtained for SiO$_x$-free Fe/MgO/$n^+$-Si and Fe/Mg (2 nm)/MgO/$n^+$-Si junctions on a $n^+$-Si substrate (see Section S9 in S.M. [17]). In addition, $\alpha = 0$ estimated from the regression analysis also supports the presence of the Fermi level pinning (The required condition cannot be satisfied without Fermi level pinning).

Based on the conditions described above, $E_C(z)$ and $n(z)$ were calculated under various $N_D$ (= $3 \times 10^{18} - 1 \times 10^{20}$ cm$^{-3}$) and $\Phi_{\text{pin}}$ (= 0 – 1.1 eV). Then, $N_{S0}^n$ was calculated from $n(z)$ through Eq. (4), and $N_D$ and $\Phi_{\text{pin}}$ were explored so that $N_{S0}^n = 2 \times 10^{12}$ cm$^{-2}$ is obtained. Here, a calculated result with $N_D = 3 \times 10^{19}$ cm$^{-3}$ and $\Phi_{\text{pin}} = 0.55$ eV is shown in Fig. 4(b), where dashed and solid black curves in the top figure represent $E_C(z)$ at $V_{GS} = 0$ and 100 V, respectively, and dashed and solid red curves in the bottom figure represent $n(z)$ at $V_{GS} = 0$ and 100 V, respectively. As shown in the top of Fig. 4(b), in the $0 < z < t_n$ range, the depletion layer formed at the surface side (the MgO/SiO$_x$ side) has a thickness of 3 – 4 nm, and the band bending of the inversion layer (the BOX layer side) has a thickness of ~2 nm. Hence, $n(z)$ is largely affected by the band bending from both sides because its length (2 - 4 nm) is comparable to the thickness ($t_n = 5$ nm) of the $n^+$-Si region along the *z* direction. In the bottom of Fig. 4(b), $n(z)$ is shown by the red curves, where electrons are depleted in $0 < z < 3$ nm by the Fermi level pinning regardless of the $V_{GS}$ application. On the other hand, electrons are present in $z > 3$ nm, and they are more accumulated in $z > 3$ nm by the $V_{GS}$ (= 100 V) application. Note that $\Phi_{\text{pin}} = 0.55$ eV agrees with a literature value ($\Phi_{\text{pin}} = 0.82 \pm 0.36$ eV [25-29]) and $N_D = 3 \times 10^{19}$ cm$^{-3}$ does not differ significantly from the designed value (~$1 \times 10^{20}$ cm$^{-3}$). From the above analysis, the smaller $N_{S0}^n$ (= $2 \times 10^{12}$ cm$^{-2}$) value can be explained by the Fermi level pinning and the slight reduction in $N_D$.

We found that $P_S$ is overestimated when large $N_{S0}^n$ is used for the analysis since $|\Delta V_{SV}|$ is roughly proportional to $P_S^2$ and $r_n \propto 1/N_{S0}^n$. When the designed $N_{S0}^n$ value ($5 \times 10^{13}$ cm$^{-2}$) is used, $P_S$ exceeds 100% and it varies greatly depending on $V_{GS}$, as shown in Fig. S12 in S.M. [17]. These results contradict our experimental results ($P_S \sim 41\%$) [15, 16] as well as the theoretical prediction, i.e., the maximum $P_S$ value is the spin polarization at the Fermi level of the ferromagnetic layer without a spin-filter insulating layer (the spin polarization at $E_F$ of Fe is 42% [31]). This fact confirms again that the realistic $N_{S0}^n$ value is needed for accurately estimating $P_S$, and that the real $N_{S0}^n$ value is decreased by the Fermi level pinning at the SiO$_x$/$n^+$-Si interface. Note that our reconstructed model presented in this study does not contradict our previous results [9, 11]: When we apply the *reconstructed* model to the *previous* experimental data, consistent $P_S$ that depends only on $V_J$ is obtained while the other parameters and characteristics are unchanged (see Section S7 in S.M. [17]).

**IV. MR ratio under various bias conditions**

Figure 5(a) shows MR ratio ($\gamma^{MR}$) plotted as a function of $V_J$, estimated by $\Delta V_{SV}/V_{DS}$ with various $I_{DS}$ ($0 < I_{DS} \leq 4$ mA) and $V_{GS}$ (= 40, 60, and 80 V). For clarity, only the positive bias region ($I_{DS} > 0$) is discussed hereafter. Figure 5(a) finds that $\gamma^{MR}$ shows a peaked feature against $V_J$ and decreases monotonically as $V_{GS}$ is increased from 40 to 80 V. The maximum $\gamma^{MR}$ value is 0.35% at $V_J = 0.43$ V ($I_{DS} = 2$ mA) and $V_{GS} = 40$ V, which is the highest MR ratio among the Si-based spin MOSFET structures [5-16], mostly owing to the achievement of the high $P_S$ value (> 40%). To understand the underlying physics further, the $V_J$ dependences on various parameters are discussed below (see also Eqs. S10(a-g) in Section S5 in S.M. [17] for more detail).

When the spin drift dominates the spin transport in the channel, $\gamma^{MR}$ can be approximately written by the following simple form (see Section S5 in S.M. [17] and Ref. [9]),

$$\gamma^{MR} = \frac{\Delta V_{SV}}{V_{DS}} \approx 2P_S^2 \frac{r^{output}}{R_T} \gamma^{LSC} \gamma^d \approx P_S^2 \frac{r^{output}}{R_J} \gamma^{LSC} \gamma^d, \quad (8)$$

$$\gamma^d = \exp\left(-\frac{L_{eff}}{\lambda_{ch}^d}\right), \quad (9)$$

$$\gamma^{LSC} = \frac{1/r_{ch}^d}{1/r_{ch}^d + 1/r_{NL}^S}, \quad (10)$$

$$\gamma^{output} = \left(\frac{1}{r_{NL}^D} + \frac{1}{r_{ch}^u}\right)^{-1}, \quad (11)$$

where $r^{output}$ is the output spin resistance, $\gamma^{LSC}$ is the local spin current ratio, $\gamma^d$ is spin transport efficiency through the Si 2D channel, $\lambda_{ch}^d$ is the down-stream spin diffusion length in the Si 2D channel, $r_{ch}^{u(d)}$ is the up(down)-stream spin resistance in Si 2D channel, and $r_{NL}^{S(D)}$ is the spin resistance of the nonlocal region on the left side (right side) of the S(D) electrode defined in Eq. (7). The nonlocal and local regions are indicated in Fig. 1(b). The local spin current ratio $\gamma^{LSC} = I_S^{local}/(P_S I_{DS})$ introduced in Ref. [9] expresses the efficiency of the spin injection into the Si 2D channel, where $P_S I_{DS}$ is the total spin current injected from the S electrode into the $n^+$-Si region and $I_S^{local}$ is the spin current that is actually injected into the Si 2D channel from the $n^+$-Si region and flows toward the D electrode.

Since the spin current is inversely proportional to the spin resistance, $\gamma^{LSC}$ is expressed by Eq. (10) (see Eq. (9) in Ref. [9] for details). As $r_{NL}^{S}$ decreases, the ratio of the spin current flow into nonlocal region increases, and thus $\gamma^{LSC}$ decreases. The output spin resistance $r^{output}$ introduced in Ref. [9] is the effective spin resistance of the whole device seen from the D electrode, and it is expressed by the parallel resistance of $r_{NL}^{D}$ and $r_{ch}^{u}$ as shown in Eq. (10). This is because the spin accumulation is formed at around the D electrode by the spin current that is transported from S electrode through the Si 2D channel, and some of them flow back into the Si 2D channel and the others diffuse into the nonlocal region on the right side of the D electrode. As $r_{NL}^{D}$ ($\propto r_n$) decreases, the spin diffusion into the nonlocal region increases, and thus $r^{output}$ decreases. On the other hand, the spin transport efficiency $\gamma^{d}$ represents the conservation rate of the spin current through the Si 2D channel and thus it depends on $I_{DS}$ and $V_{GS}$ but is independent of $r_n$, as shown in Eq. (9). In Eq. (8), $R_T$ is the total resistance consisting of $R_J$ and $R_{ch}$, i.e., $R_T = R_{ch} + 2R_J$, where $R_T = V_{DS}/I_{DS}$. The rightmost formula of Eq. (8) is derived using an approximation $R_T \cong 2R_J$ since $R_{ch}$ is only ~5% of $R_T$ in the present device. Figures 5(b), (c), and (d) show calculated $r^{output}$, $\gamma^{LSC}$, and $\gamma^{d}$ as a function of $V_J$, respectively, in which their maximum changes at each $V_{GS}$ are less than twice. These three parameters are relevant to the spin transport through the Si part composed of the $n^{+}$-Si regions at S/D electrodes and Si 2D channel, and their gradual increases with increasing $V_J$ is explained by the increase in the spin drift [20-24]. Furthermore, this study clarified that the features in $r^{output}$ and $\gamma^{LSC}$ are closely related to the band profile of the junction in Fig. 4(b), because these parameters are dominated by $r_n$ through $r_{NL}^{S(D)}$ as shown in Eqs. (11), (10) and (7), where $r_n$ is largely changed by the $V_{GS}$ application and the Fermi level pinning through Eqs. (1) and (3). The spin MOSFET examined in this study has larger $\Delta V_{SV}$ by ~20 times than the previous device reported in Ref. [9], which enables us to clarify the detailed relationship between the spin transport and band profile of the junction. In addition, the strong decrease in $r^{output}$ with increasing $V_{GS}$ in Fig. 5(b) can be explained by the reduction in $r_n$ in Eq. (1), leading to smaller $\gamma^{MR}$ at higher $V_{GS}$ in Fig. 5(a). The above features of $r^{output}$, $\gamma^{LSC}$, and $\gamma^{d}$ mean that higher $I_{DS}$ and lower $V_{GS}$ are effective to obtain higher $\Delta V_{SV}$ in Eq. (5), once the device structure is designed and prepared.

On the other hand, $P_S^2$ ($= P_{inj} \times P_{det}$) in Eq. (8) has a significant impact on $\gamma^{MR}$ since $P_S$ decreases from 50% to 21% depending on $V_J$ as shown in Fig. 4(d), leading to more than 6 times change in $P_S^2$. Our previous paper [16] revealed that $P_S$ is determined by the band structure of Fe and the band alignment between Fe and $n^+$-Si in the tunnel junction. Thus, $P_S$ is probably independent of the parameters in $n^+$-Si and Si 2D channel; $r^{\text{output}}$, $\gamma^{\text{LSC}}$, and $\gamma^d$. The last remaining parameter $2/R_T = \sim 1/R_J$ in Eq. (5) is just an electrical property, but it also has a high impact on $\gamma^{\text{MR}}$ since its maximum change in Fig. 5(e) is more than 10 times. It is noteworthy that both $1/R_J$ and $P_S$ have the strong dependence on $V_J$ as shown in Figs. 5(e) and 4(d), respectively. Hereafter, we first analyze the relationship between $\gamma^{\text{MR}}$ and $P_S$ using $P_S$–$V_J$ and $1/R_J$–$V_J$ plots, and then discuss how to realize a spin MOSFET with higher $\gamma^{\text{MR}}$, focusing on the design of the ferromagnetic tunnel junctions for the S and D electrodes. Figure 5(f) shows $\gamma^{\text{MR}}$ and $P_S$ plotted as a function of $V_J$ at $V_{\text{GS}} = 40$ V, where they are the same data shown in Figs. 5(a) and 4(d), respectively. A prominent feature is that the peak position of $\gamma^{\text{MR}}$ ($V_J \sim 0.82$ V) is higher than that of $P_S$ ($V_J \sim 0.33$ V). The peak in the $P_S$–$V_J$ plot originates from the steep increase in $P_{\text{det}}$ in a lower $V_J$ range followed by the gradual decrease in $P_{\text{Fe}}$ in a higher $V_J$ range as explained in Section III-B (see Ref. [16] for detail). On the other hand, the peak in the $\gamma^{\text{MR}}$–$V_J$ plot originates from the difference in the $V_J$ dependence between $P_S$ and $1/R_J$ ($= I_{\text{DS}}/V_J$) since $\gamma^{\text{MR}} \propto P_S^2/R_J$ as shown in Eq. (8); $P_S^2$ decreases and $1/R_J$ increases as $V_J$ increases in $V_J > 0.33$ V. Thus, the increase in $1/R_J$ in Fig. 5(e) causes the difference in peak position between $P_S$ and $\gamma^{\text{MR}}$ as $V_J$ increases. In other words, $\gamma^{\text{MR}}$ is strongly suppressed by the small $1/R_J$ (large $R_J$) at $V_J = 0.33$ V where $P_S$ takes the peak value (= 50%) in Fig. 4(d). Thus, the guideline for the enhancement of $\gamma^{\text{MR}}$ is to realize a device structure with a large $1/R_J$ at this $V_J = 0.33$ V, so that the maximum $\gamma^{\text{MR}}$ and maximum $P_S$ should be at the same $V_J$ value (= 0.33 V) in Fig. 5(f). If $1/R_J$ were constant at $1.1 \times 10^{-3}$ $\Omega^{-1}$ (which is the $1/R_J$ value at $V_J = 0.82$ V in Fig. 5(e)), the peak positions would be located at the same $V_J$ (= 0.33 V) and MR would be increased to 0.7%.

**V. Discussion**

In the present device, the $n^+$-Si regions are formed underneath the S and D electrodes to reduce parasitic resistance and obtain sufficiently large $I_{\text{DS}}$; however, $1/R_J$ in the lower $V_J$ range is significantly smaller ($1/R_J < 0.5 \times 10^{-3}$ $\Omega^{-1}$) because there is a depletion layer caused by the Fermi level pinning at the top Fe/Mg/MgO/SiO$_x$/$n^+$-Si interface (see Fig. 4(b)). Thus, a challenge lies in finding an insulating layer that allows us to remove the Fermi level pinning. Nevertheless, the use of such an insulating layer is not sufficient to obtain a large $1/R_J$ at lower $V_J$ if a high Schottky barrier is formed at the interface of the $n^+$-Si layer, when the Fermi level of the top ferromagnetic metal layer is located in the middle of the band gap of Si. This issue can be solved when the work function of the top metal layer is reduced using an insertion layer, such as Mg, between the ferromagnetic and insulating layers. This method was demonstrated in our previous paper, in which a thin Mg insertion layer with a sufficient thickness (~1 nm) can reduce the resistance of a Fe/Mg/SiO$_x$N$_y$/$n^+$-Si junction while the Fermi level pinning is removed by the SiO$_x$N$_y$ layer [32, 33]. When this junction is used for the S/D of the spin MOSFET, the peak position of $P_S$ (~16%) and $\gamma^{MR}$ will be located at closer $V_J$ values ($V_J$ ~ 0.1 and 0.3 V, respectively) owing to the higher $1/R_J$ (~ $1.5 \times 10^{-1}$ $\Omega^{-1}$ [32]), and thus higher MR (~ 2%) is expected even though the peak $P_S$ value is lower. The method of this work-function engineering also brings about another large advantage, i.e., the Si region in the tunnel junction at the S/D electrodes can have a low doping concentration and an almost flat band profile under any bias conditions. In this situation, $r_n$ becomes larger and insensitive to $V_J$, and both $r^{output}$ and $\gamma^{LSC}$ significantly increase in the lower $V_J$ range. Note that the increase in $r_n$ changes $R_T$ little since the $n^+$-Si regions underneath the S and D have much lower resistance than the Si 2D channel. Therefore, all the enhancements of $P_S$, $1/R_J$, $r^{output}$, and $\gamma^{LSC}$ contribute to a significant enhancement of $\gamma^{MR}$, as shown in Eq. (8). Further enhancement is also expected when a top gate-type device structure is used because unwanted spin diffusion into the nonlocal region can be suppressed, as theoretically demonstrated in our previous paper [9].

**VI. Conclusion**

We studied the spin MOSFET with Fe/MgO/SiO$_x$/$n^+$-Si ferromagnetic S/D junctions and achieved the maximum $\gamma^{MR}$ value of 0.35% at 4 K, that is the highest value reported so far owing to the high $P_S$ of the junction structure. Our detailed analyses on the junction properties revealed that the band profile of the $n^+$-Si region has a large influence on the estimation of $P_S$, which is substantially changed by $V_J$ and $V_{GS}$ since the 5-nm-thick $n^+$-Si region has band bending at the front and back sides due to the Fermi level pinning at the SiO$_x$/Si interface and the Si 2D channel formation near the Si/BOX interface, respectively. The spin transport model was reconstructed by taking into account the electron distribution and band profile of the $n^+$-Si region in the ferromagnetic S/D junction, which leads to the accurate estimation of $P_S$. The estimated $P_S$ value shows a peaked feature against $V_J$ with the maximum value of ~50%. The obtained $\gamma^{MR}$ also shows a peaked feature (maximum 0.35 %) against $V_J$. These $P_S$ and $\gamma^{MR}$ values are the highest ever reported in spin-MOSFETs. However, the $V_J$ value giving the maximum $\gamma^{MR}$ is higher by ~ 0.4 V than $V_J$ giving the maximum $P_S$ because the $V_J$-dependent $1/R_J$ is smaller in the lower $V_J$ range. This means that the present device does not use the full potential of $P_S$ for maximizing $\gamma^{MR}$, since $\gamma^{MR}$ ($\propto P_S^2/R_J$) is strongly suppressed by the small $1/R_J$ at $V_J$ that gives the maximum $P_S = 50\%$. Based on this result, to obtain higher $\gamma^{MR}$, we need to realize the condition that the maximum $\gamma^{MR}$ and maximum $P_S$ are located at the same $V_J$ value. To achieve this goal, ferromagnetic tunnel junctions at the S/D electrodes are required to have the following two characteristics at the same time: High $P_S$ and low junction resistance-area product through the removal of Fermi level pinning.

## Acknowledgments

This work was partly supported by Grants-in-Aid for Scientific Research (20H05650, 23H00177, 23K17324, 25H00840), CREST Program (JPMJCR1777) of Japan Science and Technology Agency, and Spintronics Research Network of Japan (Spin-RNJ).

## Data availability

The data that support the findings of this article are openly available [35].

## References

[1] S. Sugahara and M. Tanaka, Appl. Phys. Lett. **84**, 13 (2004).
[2] S. Sugahara and M. Tanaka, ACM Transactions on Storage **2**, 197 (2006).
[3] M. Tanaka and S. Sugahara, IEEE Trans. Electron Devices **54**, 961 (2007).
[4] H.-J. Jang and I. Appelbaum, Phys. Rev. Lett. **103**, 117202 (2009).
[5] R. Nakane, T. Harada, K. Sugiura, and M. Tanaka, Jpn. J. Appl. Phys. **49**, 113001 (2010).
[6] T. Sasaki, Y. Ando, M. Kameno, T. Tahara, H. Koike, T. Oikawa, T. Suzuki, and M. Shiraishi, Phys. Rev. Applied **2**, 034005 (2014).
[7] T. Tahara, H. Koike, M. Kameno, S, Sasaki, Y. Ando, K. Tanaka, S. Miwa, Y. Suzuki, and M. Shiraishi, Appl. Phys. Express **8**, 113004 (2015).
[8] S. Sato, M. Ichihara, M. Tanaka, and R. Nakane, Phys. Rev. B **99**, 165301 (2019).
[9] S. Sato, M. Tanaka, and R. Nakane, Phys. Rev. B **102**, 035305 (2020).
[10] R. Nakane, S. Sato, and M. Tanaka, IEEE J. Electron Devices Soc. **8**, 807 (2020).
[11] S. Sato, M. Tanaka, and R. Nakane, Phys. Rev. Appl. **18**, 064071 (2022).
[12] M. Tanaka, Jpn. J. Appl. Phys. **60**, 010101 (2021).
[13] H. Koike, S. Lee, R. Ohshima, E. Shigematsu, M. Goto, S. Miwa, Y. Suzuki, T. Sasaki, Y. Ando and M. Shiraishi, Applied Physics Express **13**, 083002 (2020). They have reported 1.4% of MR ratio at 300 K, but their device is not a spin MOSFET structure.
[14] T. Endo, S. Tsuruoka, Y. Tadano, S. Kaneta-Takada, Y. Seki, M. Kobayashi, L. D. Anh, M. Seki, H. Tabata, M. Tanaka, S. Ohya, Adv. Mater. **35**, 2300110 (2023). They have reported ~140% of MR ratio at 3 K on STO-based lateral devices with a back-gate electrode.

[15] R. Nakane, M. Ichihara, S. Sato, and M. Tanaka, Phys. Rev. Mater. **3**, 024411 (2019).

[16] B. Yu, S. Sato, M. Tanaka, and R. Nakane, Phys. Rev. B **110**, 205302 (2024).

[17] See Supplemental Material [url] for S1. Electrical and spin-related parameters used for our analysis, S2. Spin-valve signals at $V_{GS}$ = 60 and 80 V, S3. Self-consistent calculation of the conduction band profile $E_C(z)$ and electron density $n(z)$ distribution in the $n^+$-Si region, S4. Detailed derivation of Eq. (1) in the main manuscript, S5. Reconstructed 2T Hanle signal expression, S6. Comparison of the $P_S$–$V_J$ characteristics with our previous experimental and theoretical results, S7. Verification of our previous results, S8. 2T Spin-valve signal analysis in the Spin MOSFET with $L_{ch}$ = 0.85 μm, S9. Current – voltage ($I$ – $V$) characteristics of MgO-based tunnel junctions, S10. Analysis of the spin-valve signals using the previous model, S11 Detailed expression of the regression analysis, which includes Refs. [36-38].

[18] F. J. Jedema, H. B. Heersche, A. T. Filip, J. J. A. Baselmans, and B. J. van Wees, Nature (London) **416**, 713 (2002).

[19] F. J. Jedema, M. V. Costache, H. B. Heersche, J. J. A. Baselmans, and B. J. van Wees, Appl. Phys. Lett. **81**, 5162 (2002).

[20] Z. G. Yu and M. E. Flatté, Phys. Rev. B **66**, 235302 (2002).

[21] M. Kameno, Y. Ando, T. Shinjo, H. Koike, T. Sasaki, T. Oikawa, T. Suzuki, and M. Shiraishi, Appl. Phys. Lett. **104**, 092409 (2014).

[22] T. Miyakawa, T. Akiho, Y. Ebina, M. Yamamoto, and T. Uemura, Appl. Phys. Exp. **9** (2016).

[23] A. Tiwari, T. Inokuchi, M. Ishikawa, H. Sugiyama, N. Tezuka, and Y. Saito, Jpn. J. Appl. Phys. **56**, 4S (2017).

[24] Y. Takamura, T. Akushichi, Y. Shuto, and S. Sugahara, J. Appl. Phys. **117**, 17D919 (2015).

[25] S. M. Sze and M. K. Lee, *Physics of Semiconductor Devices*, 3rd ed. (John Wiley and Sons, New York, 2007). P.144.

[26] A. M. Cowley and S. M. Sze, J. Appl. Phys. **36**, 3212 (1965).

[27] T. Nishimura, K. Kita, and A. Toriumi, Appl. Phys. Lett. **91**, 123123 (2007).

[28] R. T. Tung, Phys. Rev. Lett. **84**, 6078 (2000).

[29] W. Mönch, J. Vac.Technol. B **17** (1999).

[30] S. Sato, R. Nakane, T. Hada, and M. Tanaka, Phys. Rev. B **96,** 235204 (2017).

[31] R. Meservey and P.M. Tedrow, Phys. Rep. **238**, 173 (1994).

[32] R. Nakane, T. Hada, S. Sato, and M. Tanaka, Appl. Phys. Lett. **112**, 182404 (2018).

[33] D. Connelly, C. Faulkner, P. A. Clifton, and D. E. Grupp, Appl. Phys. Lett. **88**, 012105 (2006).

[34] T. Valet and A. Fert, Phys. Rev. B **48**, 7099, (1993).

[35] S. Sato, M. Tanaka, and R. Nakane, Zenodo (2025), doi: https://doi.org/10.5281/zenodo.16311236

[36] S. O. Valenzuela, D. J. Monsma, C. M. Marcus, V. Narayanamurti, and M. Tinkham, Spin polarized tunneling at finite bias, Phys. Rev. Lett. **94**, 196601 (2005).

[37] J. Callaway and C. S. Wang, Phys. Rev. B **16**, 2095 (1977).

[38] S. Zhang, P. M. Levy, A. C. Marley, and S. S. P. Parkin, Phys. Rev. Lett. **79**, 3744 (1997).

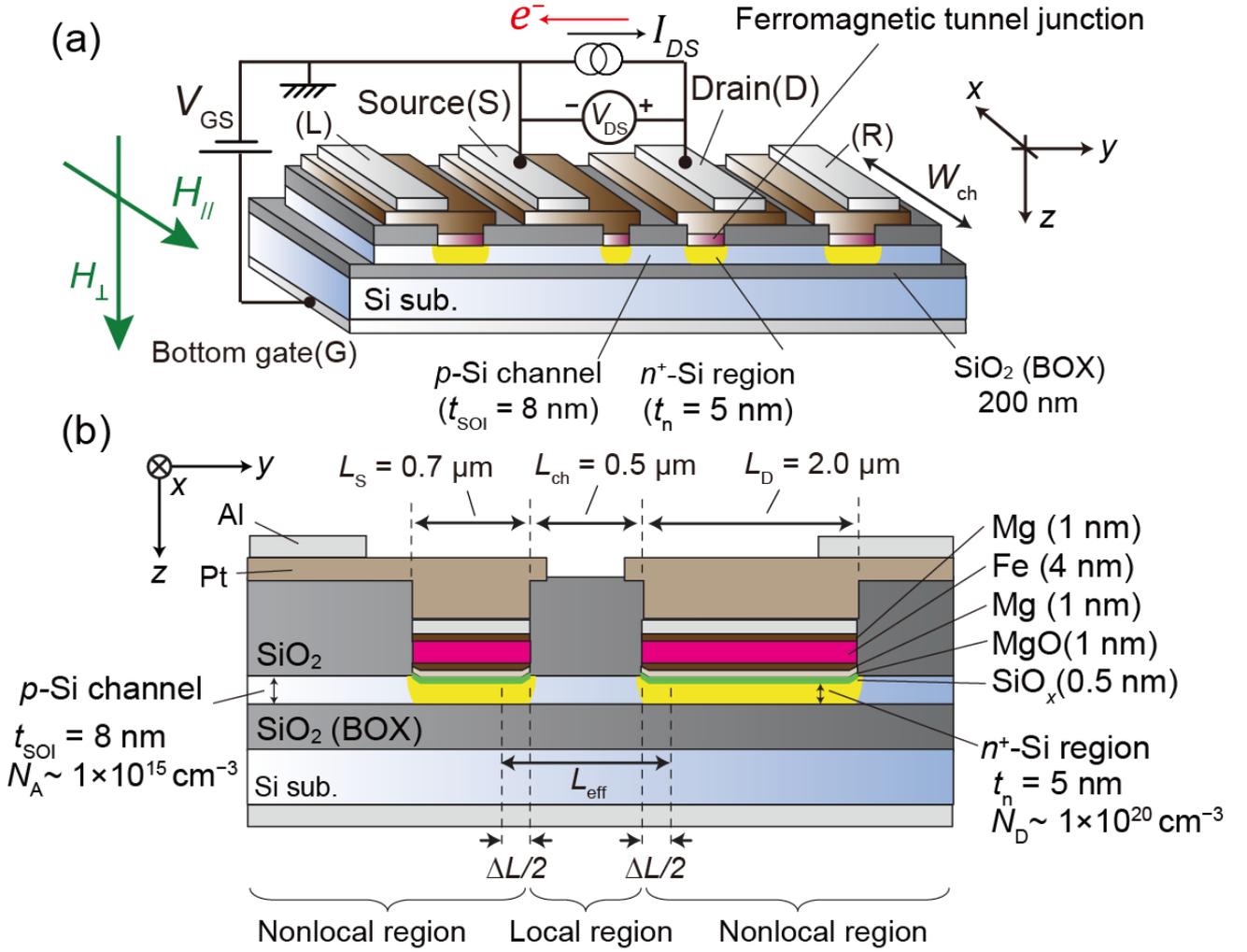

**Figure 1** (a) Schematic illustration of a spin MOSFET structure on a silicon-on-insulator (SOI) substrate, where the thickness of a *p*-Si channel is $t_{SOI} = 8$ nm and the source (S) and drain (D) have a Fe (4 nm)/Mg (2 nm)/MgO(1 nm)/SiO$_x$(0.2 nm)/$n^+$-Si($t_n = 5$ nm) ferromagnetic tunnel junction. The *p*-Si channel has a boron doping concentration $N_A$ of $1 \times 10^{15}$ cm$^{-3}$ and the $n^+$-Si regions represented by yellow areas have a phosphorus donor doping concentration $N_D$ of is $1 \times 10^{20}$ cm$^{-3}$. Our two-terminal (2T) measurement setup to obtain spin signals is also shown, where the voltage between the D and S electrodes $V_{DS}$ are measured by a voltage meter while a constant drain-source current $I_{DS}$ and a constant gate-source voltage $V_{GS}$ are applied and an external magnetic field is swept along the in-plane ($H_{//}$) or perpendicular ($H_\perp$) direction. The Cartesian coordinate is defined as follows: the *x* and *y* axes are parallel to the channel width and length direction, respectively, and the *z* axis is perpendicular to the substrate plane. (b) Magnified side view around the center of the device, where the S and D have a (from top to bottom) Al(15 nm)/Mg(1 nm)/Fe(4 nm)/Mg(1 nm)/MgO(1 nm)/SiO$_x$(0.2 nm)/$n^+$-Si ferromagnetic tunnel junction, the channel length $L_{ch}$ defined by the gap between the S and D is 0.5 µm, and the lengths of the S ($L_S$) and the D ($L_D$) electrodes are 0.7 and 2.0 µm, respectively. The effective channel length $L_{eff}$ is estimated by an electrical measurement, which is defined by $L_{eff} = L_{ch} + \Delta L$. Local and nonlocal regions are defined, where the local region corresponds to the Si 2D channel between the S and D electrode, and the nonlocal regions are the outside regions including the $n^+$-Si regions.

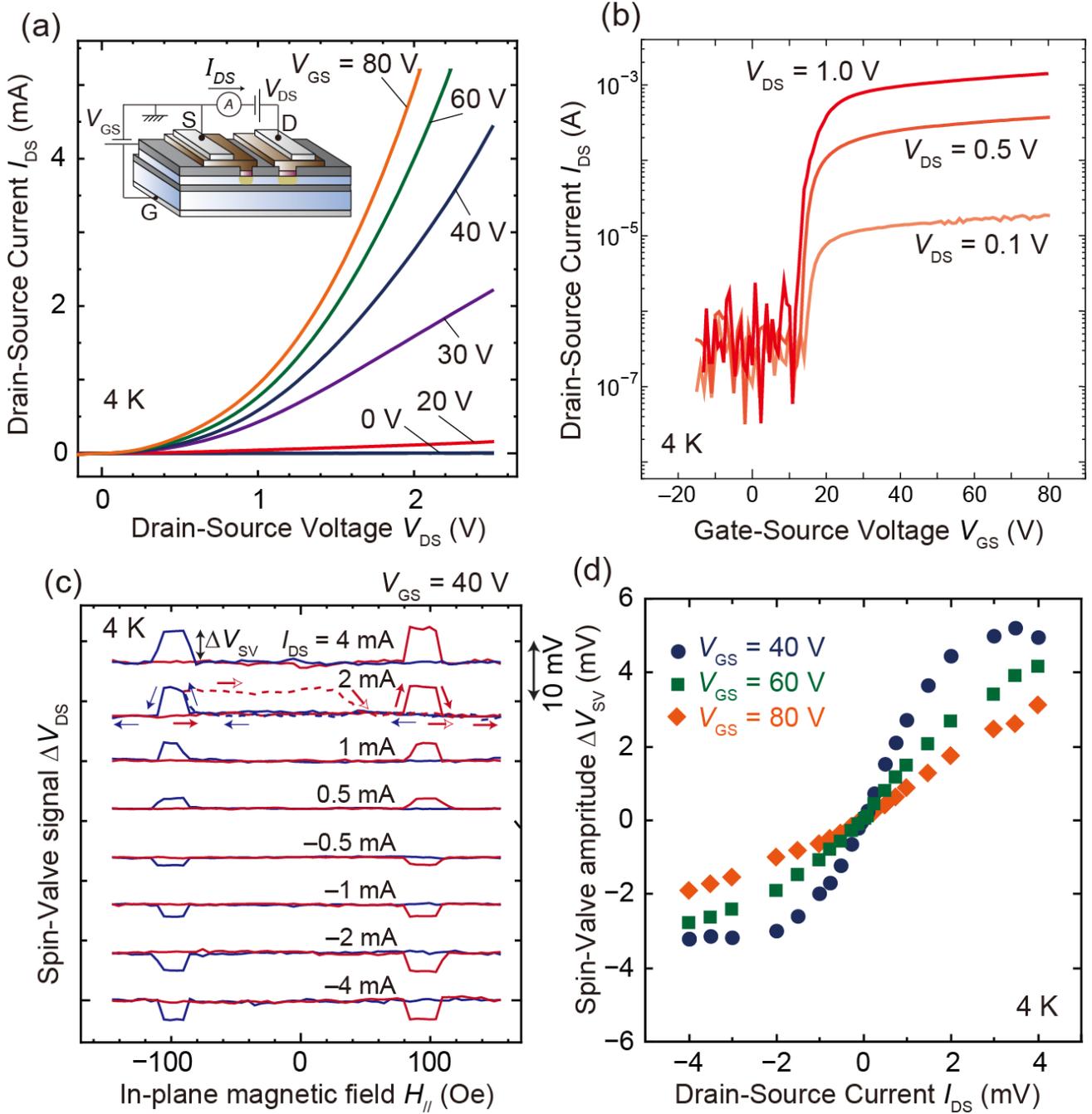

**Figure 2** (a) $I_{DS}$–$V_{DS}$ characteristics measured at 4 K for the spin MOSFET when the measurement setup in the figure is used and $V_{GS}$ is varied from 0 to 80 V in steps of 20 V. (b) $I_{DS}$–$V_{GS}$ characteristics measured at 4 K when a constant $V_{DS}$ (= 0.1, 0.5, or 1.0 V) is applied. (c) Spin-valve signals $\Delta V_{DS}$ measured at 4 K with various $I_{DS}$ (= −4 – 4 mA) and $V_{GS}$ = 40 V, where red and blue curves represent the signals measured with negative and positive sweeping directions of $H_{//}$, respectively. The amplitude of the spin-valve signal $\Delta V_{SV}$ is defined, which is the maximum change between the P and AP magnetization configurations. (d) $\Delta V_{SV}$ plotted as a function of $I_{DS}$, where the measurement temperature is 4 K and blue circles, green squares, and orange diamonds are $\Delta V_{SV}$ data at $V_{GS}$ = 40, 60, and 80 V, respectively.

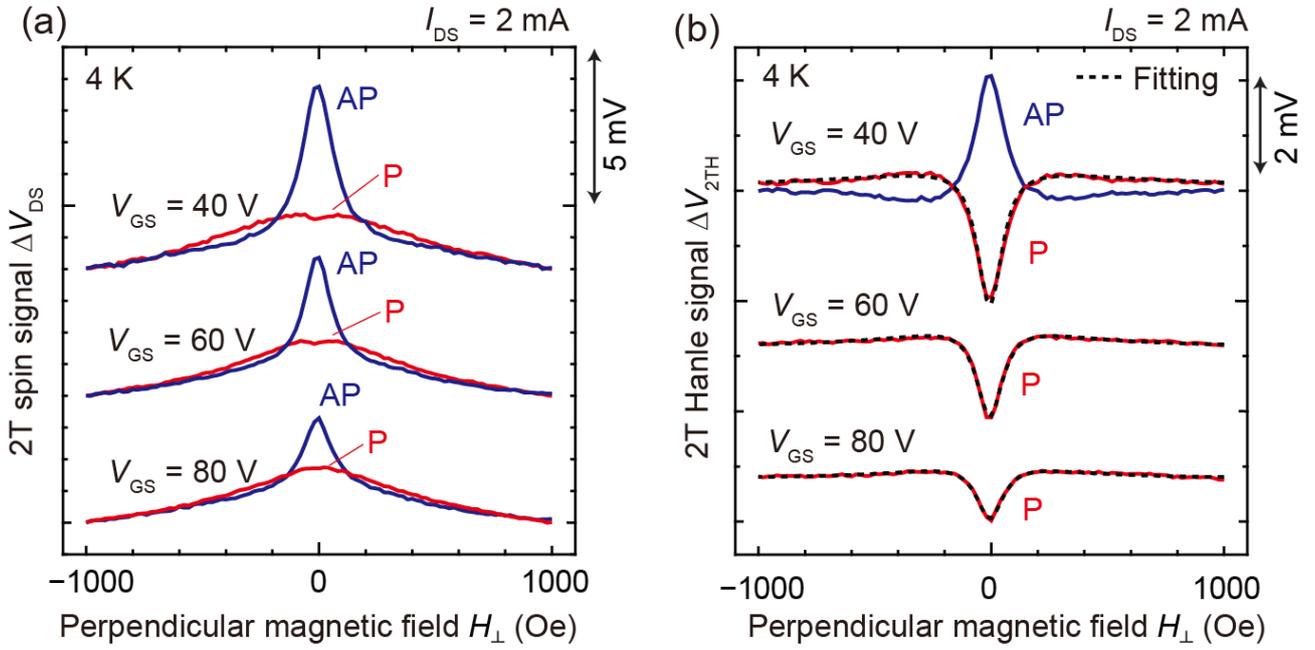

**Figure 3** (a) Two-terminal (2T) spin signals $\Delta V_{DS}^{P/AP}(H_\perp)$ measured at 4 K with $I_{DS}$ = 2 mA and $V_{GS}$ = 40, 60, and 80 V when a perpendicular magnetic field $H_\perp$ was swept, where red and blue curves represent the signals in the parallel (P) and antiparallel (AP) magnetization configurations. (b) 2T Hanle precession signals extracted from the spin signals in (a) using a formula $\Delta V_{2TH}^{AP(P)}(H_\perp) = +(-)[\Delta V_{DS}^{AP}(H_\perp) - \Delta V_{DS}^{P}(H_\perp)]/2$, where red and blue curves are the experimental signals in the AP (+) and P (−) magnetization configurations, respectively, and black dashed curves are the fitting curves with Eq. (5). $\Delta V_{2TH}^{AP}(H_\perp)$ for $V_{GS}$ = 60 and 80 V are not shown.

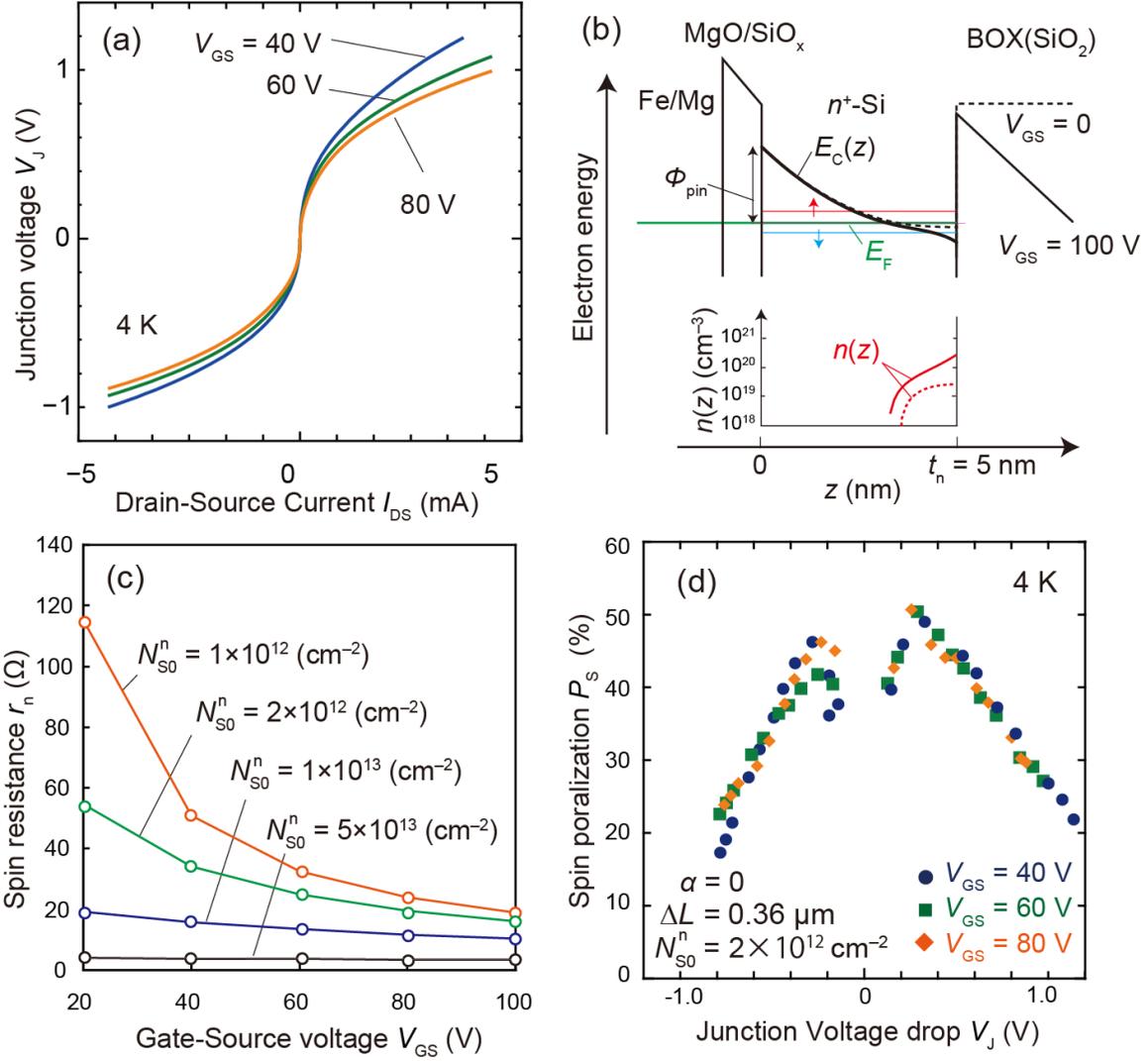

**Figure 4** (a) Junction voltage $V_J$ plotted as a function of $I_{DS}$ at $V_{GS}$ = 40, 60, and 80 V, which was estimated from a formula $V_J = (V_{DS} - I_{DS} \times R_{ch})/2$ using the $I_{DS} - V_{DS}$ data in Fig. 2(a) and the channel resistance $R_{ch}$ estimated with a Hall-bar-type MOSFET (see Section S1 in S.M.). (b) (Top) One-dimensional potential profile of the conduction band minimum $E_C(z)$ along the vertical $z$ direction in the $n^+$-Si region where $N_D = 3 \times 10^{19}$ cm$^{-3}$, $\Phi_{pin} = 0.55$ eV (in the mid-gap of Si), and $z = 0$ is defined at the SiO$_x$/Si interface, black-dashed and black-solid curves in $0 < z < 5$ nm are $E_C(z)$ at $V_{GS} - V_{th} = 0$ V and 100 V, respectively, red and blue horizontal lines are the chemical potentials for up- and down-spin electrons, and a green solid line is the Fermi level $E_F$. The potentials in $z < 0$ and $z > 5$ nm are $E_C(z)$ of the MgO/SiO$_x$ and BOX layer, respectively. (Bottom) Electron density distribution $n(z)$ (cm$^{-3}$) corresponding to the top figure, where red-dashed and red-solid curves in $0 < z < 5$ nm are $n(z)$ at $V_{GS} - V_{th} = 0$ V and 100 V, respectively. (c) Spin resistance $r_n$ (Ω) under various $V_{GS}$ (= 20 – 100 V) and various $N_{S0}$ values, which were calculated using Eq. (1) with $V_J = 0$. Black, blue, green, and orange dots are calculated for $N_{S0}^n$ = 5×10$^{13}$, 1×10$^{13}$, 2×10$^{12}$, and 1×10$^{12}$ cm$^{-2}$, respectively. (d) $P_S$ plotted as a function of $V_J$, estimated from $\Delta V_{SV}$ in Fig. 2(d) using $\Delta L$ = 0.36 μm, $N_{S0}^n$ = 2×10$^{12}$ cm$^{-2}$, and $\alpha = 0$. Blue circles, green squares, and orange diamonds are $P_S$ values at $V_{GS}$ = 40, 60, and 80 V, respectively.

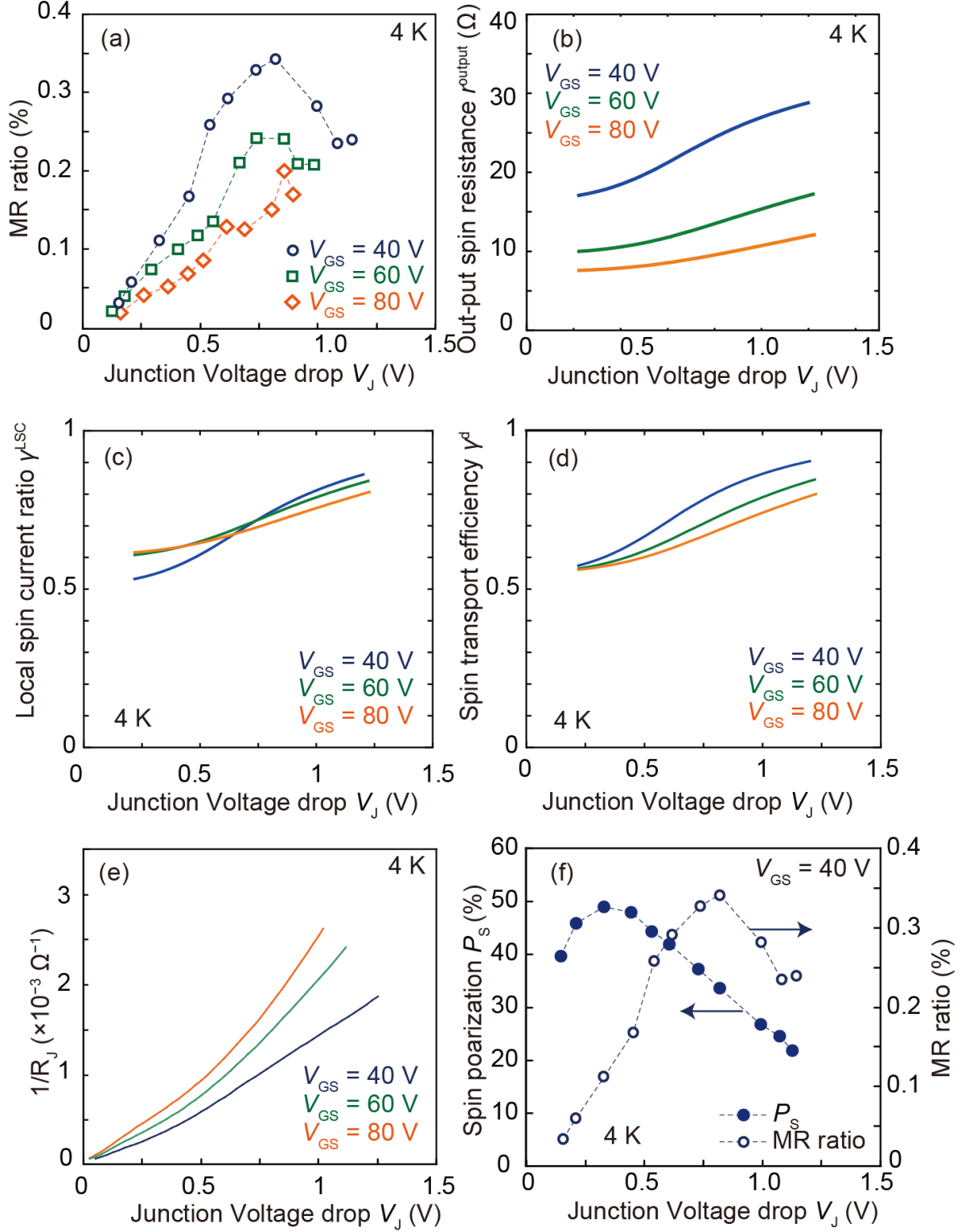

**Figure 5** (a) MR ratio $\gamma^{MR} = \Delta V_{SV}/V_{DS}$ estimated from $\Delta V_{SV}$ in Fig. 2(d), where blue circles, green squares, and orange diamonds are those at $V_{GS}$ = 40, 60, and 80 V, respectively. (b)–(e) Output resistance $r_{output}$, local spin current ratio $\gamma^{LSC}$, spin transport efficiency $\gamma^d$, and inverse of the junction resistance $1/R_J$ calculated using Eqs. (8)–(11) and $R_J = V_J/I_{SD}$, respectively. Blue, green, and orange curves are those at $V_{GS}$ = 40, 60, and 80 V, respectively. (f) $\gamma^{MR}$ (open blue circles) and $P_S$ (filled blue circles) at $V_{GS}$ = 40 V which are the same as in Fig. 5(a) and Fig. 4(d), respectively.